# Polychromatic neutron phase contrast imaging of weakly absorbing samples enabled by phase retrieval

**Maja Østergaard[a1], Estrid Buhl Naver[b1], Anders Kaestner[c], Peter K. Willendrup[de], Annemarie Brüel[f], Henning Osholm Sørensen[dg], Jesper Skovhus Thomsen[f], Søren Schmidt[e], Henning Friis Poulsen[d], Luise Theil Kuhn[b*] and Henrik Birkedal[a*]**

[a]Department of Chemistry and iNANO, Aarhus University, Gustav Wieds Vej 14, Aarhus, Denmark

[b]Department of Energy Conversion and Storage, Technical University of Denmark, Fysikvej 310, Kongens Lyngby, Denmark

[c]Laboratory for Neutron Scattering and Imaging, Paul Scherrer Institute, Villigen, Switzerland

[d]Department of Physics, Technical University of Denmark, Fysikvej 307, Kongens Lyngby, Denmark

[e]European Spallation Source ERIC, P.O. Box 176, Lund, Sweden

[f]Department of Biomedicine, Aarhus University, Wilhelm Meyers Allé 3, Aarhus, Denmark

[g]Xnovo Technology ApS, Galoche Allé 15, Køge Denmark

Correspondence email: luku@dtu.dk; hbirkedal@chem.au.dk

[1]These authors contributed equally

**Synopsis** Neutron imaging enhanced by retrieval of propagation-based phase contrast is described.

**Abstract** We demonstrate the use of a phase retrieval technique for propagation-based phase contrast neutron imaging with a polychromatic beam. This enables imaging samples with low absorption contrast and/or improving the signal-to-noise ratio to facilitate e.g. time resolved measurements. A metal sample, designed to be close to a pure phase object, and a bone sample with canals partially filled with $D_2O$ were used for demonstrating the technique. These samples were imaged with a polychromatic neutron beam followed by phase retrieval. For both samples the signal-to-noise ratio were significantly improved and in case of the bone sample, the phase retrieval allowed for separation of bone and $D_2O$, which is important for example for *in situ* flow experiments. The use of deuteration-contrast avoids the use of chemical contrast enhancement and makes neutron imaging an interesting complementary method to X-ray imaging of bone.

## 1. Introduction

There is a continued need for improved 3D imaging methods to analyse complex multi-length-scale structures of a plethora of technologically or biologically important materials. In this connection, neutron imaging is of interest due to the good penetration capabilities, the non-continuous dependence of interaction cross sections with atomic number and the possibilities for adjusting material contrast by controlling isotope compositions. The primary principle of image formation with neutron imaging is attenuation contrast from absorption. This is not always the best way to provide contrast, especially for samples consisting of materials with very similar absorption cross sections or very low absorption. For this case, it is useful to apply the principle of phase contrast that harnesses the real part of the refractive index. Phase contrast imaging can be performed in several ways, using various kinds of interferometry (Pushin *et al.*, 2017; Strobl *et al.*, 2019) or propagation-based phase contrast (Fiori *et al.*, 2006; Paganin *et al.*, 2019). The benefit of the latter is that no gratings or other specialised elements are needed to measure the phase shift. Therefore, propagation-based phase contrast imaging has evolved to a very powerful tool in X-ray imaging (Alloo *et al.*, 2022; Bidola *et al.*, 2017; Wieland *et al.*, 2021; Yu *et al.*, 2021) but has only been investigated for neutrons in a very few cases (Allman *et al.*, 2000; Jacobson *et al.*, 2004; Lehmann *et al.*, 2005; McMahon *et al.*, 2003).

When neutrons pass through a sample, they are refracted due to local variations in the refractive index. Features in the sample with different refractive indices will act as neutron lenses that either focus or diverge the neutrons from that point in the sample. These changes of the neutron directions are small and thus difficult to observe at short distances between sample and detector but they do become progressively easier to detect when increasing the distance between sample and detector. The build-up of phase contrast requires a coherent beam, meaning that the beam divergence must be smaller than the changes caused by refractive features. The propagation-based phase contrast imaging only gives relative information about the sample, which means that a phase-retrieval technique is required in order to obtain the projected density of the sample (Paganin *et al.*, 2019).

Here we explore propagation-based phase contrast imaging first in a model sample with very low absorption and secondly in bone, which is a hierarchically structured material.

Bone is replete with blood vessels and cells situated in lacunae interconnected by canaliculi only a few hundred nm in diameter (Wittig *et al.*, 2022). Together, these form a vast fluid-containing network. Transport of fluid in bone is very important since the fluid contains nutrients, signal molecules, and ions essential for the entire body (Cowin & Cardoso, 2015). In addition, liquid transport is proposed to be the main mechanism of stress sensing in bone, suggesting that the osteocytes sense changes in shear liquid flow through the canaliculi (van Tol *et al.*, 2020; Robling & Bonewald, 2020; Burger & Klein-Nulend, 1999). Our understanding of liquid transport through the complex porous network of large canals (mean Harversian canal diameter 85 µm (in iliac crest) (Busse *et al.*, 2013)), and smaller channels – all the way down to the cellular level – remains incomplete. The potential of neutron imaging to afford insights into liquid transport in bone is thus highly interesting, since neutron imaging enables contrast variation by deuteration. The field of neutron imaging of bone tissue is developing (Törnquist *et al.*, 2020; Guillaume *et al.*, 2021), but methods providing improved signal to noise and/or faster measurements remain in high demand.

Neutron propagation-based phase contrast imaging has been demonstrated for near-monochromatic neutron beams (Paganin *et al.*, 2019) but being able to harness the higher flux accessible with pink or even white neutron beams, see for example (McMahon *et al.*, 2003), would increase the efficiency and applicability of the method significantly, which is indeed the aim of the current contribution.

## 2. Experimental

### 2.1. Metal sample

To validate the use of phase retrieval of propagation-based phase contrast neutron imaging with a polychromatic neutron beam a sample was designed to have a good neutron phase contrast signal but weak neutron absorption contrast. The sample was made of Al and Zr sheets (coherent scattering cross section and thermal neutron absorption cross sections of 6.44 barn and 0.185 barn, respectively, for Zr and of 1.495 and 0.231 barn for Al) with thicknesses of 10 μm and 25 μm, respectively, such that various thicknesses were obtained (Figure 1a and 1b). The foils were cut into three pieces each of width 4 mm and heights ranging between 8 and 11 mm and assembled in a staircase configuration as shown in Figure 1b.

### 2.2. Phase retrieval

For neutron absorption imaging the projection images are formed according to the Beer-Lambert attenuation law. In phase contrast imaging this is not sufficient to describe the full contrast mechanism. Instead the projection $I(\mathbf{r}_\perp, z = \varDelta)$ at sample-detector distance $z = \Delta$ in the Fresnel regime and as a function of two-dimensional positional coordinates in the plane $\mathbf{r}_\perp = (x, y)$, can be described (Paganin *et al.*, 2019):

$$\frac{I(\mathbf{r}_\perp, z = \varDelta)}{I_0} = \left(1 - \frac{b\lambda^2\Delta}{2\pi\sigma}\nabla_\perp^2\right)\exp[-\sigma\rho_\perp(\mathbf{r}_\perp)]. \tag{1}$$

Here $I_0$ is the beam intensity without sample, $b$ is the bound coherent scattering length, $\lambda$ is the wavelength of the neutron beam, $\sigma$ is the total neutron cross section, and $\rho_\perp(\mathbf{r}_\perp)$ is the number density of atoms.

Phase-retrieval from measurements using a single sample to detector distance developed for use in X-ray propagation-based phase contrast by Paganin and co-workers (Paganin *et al.*, 2002) is sometimes referred to as Paganin filtering. It has recently been adapted for use in neutron phase contrast imaging (Paganin *et al.*, 2019).

The goal of phase retrieval is to obtain the number density $\rho_\perp(\mathbf{r}_\perp)$ of atoms of a single-material sample given the propagation-based phase contrast image $I(\mathbf{r}_\perp, z)$. Paganin *et al.* (2019) showed that this could be obtained by

$$\rho_\perp(\mathbf{r}_\perp) = -\frac{1}{\sigma}\ln\left(\mathcal{F}^{-1}\left\{\frac{\mathcal{F}\left[\frac{I(\mathbf{r}_\perp, z=\Delta)}{I_0}\right]}{1+\tau(k_x^2+k_y^2)}\right\}\right). \quad (2)$$

Here, $\mathcal{F}$ and $\mathcal{F}^{-1}$ denote the spatial Fourier and inverse Fourier transforms, $(k_x, k_y)$ are Fourier space spatial frequencies corresponding to $(x, y)$, and $\tau$ is

$$\tau = \frac{\lambda^2 b \Delta}{2\pi\sigma} - \frac{(\Theta\Delta)^2}{8}, \quad (3)$$

where $\Theta$ is the beam divergence. The core of this expression is the Fourier-space low-pass filter $1/[1+\tau(k_x^2+k_y^2)]$, which depends solely on the parameter $\tau$ that determines the strength of the phase contrast signal. The result of Eq. 2 is the number density of atoms for a given volume in the sample under the assumption that the sample consists of a homogeneous material.

It is important that $\tau > 0$, to avoid the denominator in Eq. 2 being zero. This is achieved primarily through collimation of the neutron beam. From Eq. 3 we get the collimation condition

$$\Theta < 2\lambda\sqrt{\frac{b}{\pi\sigma\Delta}}. \quad (4)$$

This presents a trade-off between having a divergent beam with higher neutron intensity, and thus better statistics, but worse phase contrast, compared to having a more coherent beam with reduced signal to noise but a better phase contrast signal. The divergence in the experiments reported herein is described below.

Eq. 2 is only valid for a monochromatic beam. For a polychromatic, pink, or white beam, the wavelength dependence should in principle be treated explicitly, which would require an energy sensitive detector. Paganin *et al.* (2019) derived an approximate treatment employing effective spectrally averaged quantities as shown in Eq. 5.

$$\rho_\perp(\mathbf{r}_\perp) = -\frac{1}{\sigma_{av}} \ln\left(\mathcal{F}^{-1}\left\{\frac{\mathcal{F}[I_{av}(\mathbf{r}_\perp, z = \Delta)]}{1 + \frac{(\sigma\tau)_{av}}{\sigma_{av}}(k_x^2 + k_y^2)}\right\}\right), \quad (5)$$

where the spectrally-averaged quantities $I_{av}$, $\sigma_{av}$, and $(\sigma\tau)_{av}$ are defined as

$$I_{av}(\mathbf{r}_\perp, z = \Delta) \equiv \frac{\overline{I_E(\mathbf{r}_\perp, z = \Delta)}}{\overline{I_{0,E}}}, \quad (6a)$$

$$\sigma_{av} \equiv \frac{\overline{\sigma_E I_{0,E}}}{\overline{I_{0,E}}}, \quad (6b)$$

$$(\sigma\tau)_{av} \equiv \frac{\overline{\sigma_E \tau_E I_{0,E}}}{\overline{I_{0,E}}}. \quad (6c)$$

Here, $I_{0,E}$ is the energy spectrum of the emitted neutrons. We used this formulation for all phase retrieval calculations in the present work.

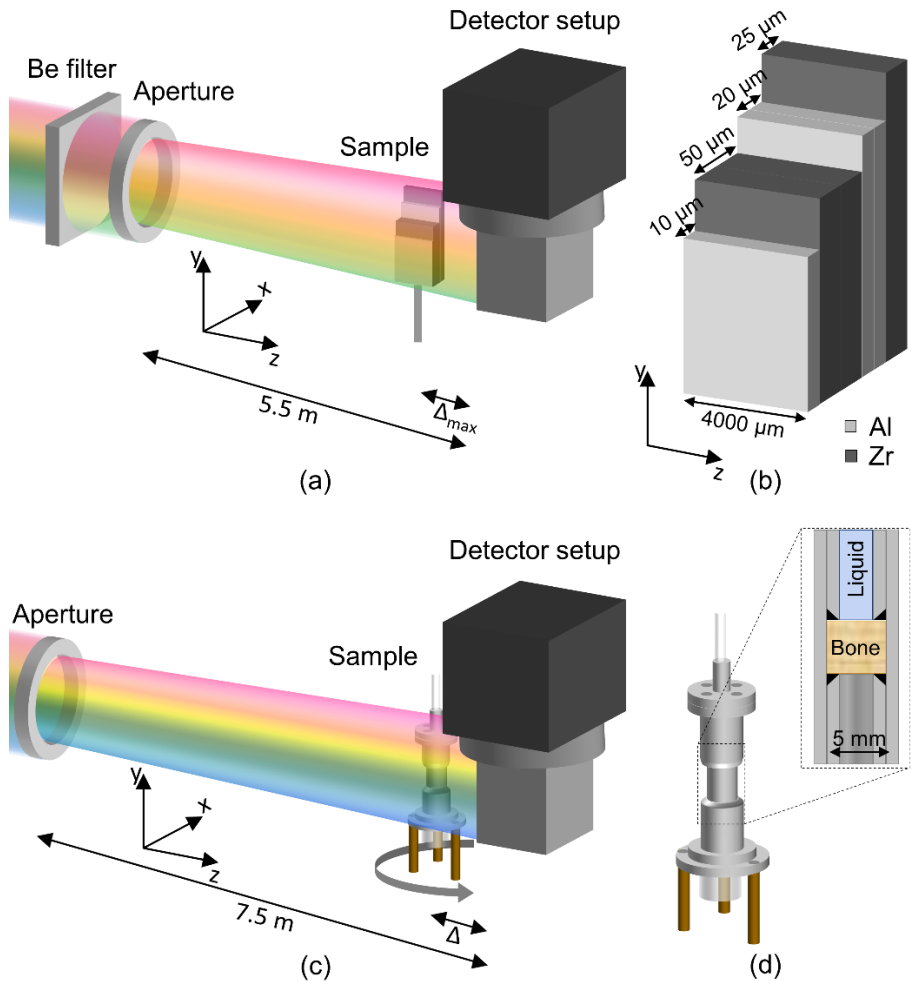

**Figure 1** Experimental setup for neutron propagation-based phase contrast measurements with either a pink or a white beam. (a) Sketch of the experimental setup for a high phase/low absorption contrast metal sample and a pink beam showing the position and angle of the sample relative to the detector and aperture. Coordinate system referring to the phase retrieval algorithm. (b) Sketch of the metal foil sample consisting of Al (light grey) and Zr (dark grey). (c) Sketch of setup for white beam imaging of bone showing the position of the sample relative to aperture and detector. (d) Sketch of the sample holder, which could be connected to a pump at the top in order to force liquid through the bone sample that is placed in the middle. A reservoir is present in the bottom to collect liquid. The inset shows a sketch of a vertical cut through sample holder, with the sample in the middle packed in between O-rings and metal pipes, holding it in place. The top pipe contains liquid while the bottom pipe is empty.

### 2.3. Neutron imaging with pink beam of metal test sample

Neutron imaging measurements on the metal test sample were performed at the Imaging with Cold Neutrons beamline (ICON) at the Paul Scherrer Institute (Kaestner *et al.*, 2011). The beam was modified with a Be-filter to supress neutrons with wavelengths below 4 Å followed by a 10 mm circular aperture to further increase spatial coherence, Figure 1a. This led to a wavelength range of 4–9 Å, with a weighted mean wavelength of 5.35 Å. The detector was a CCD camera with a 20 µm Gadox scintillator and a pixel size of 13.5×13.5 µm$^2$. The flux incident on the sample was calculated from the open beam measurements to be 4.1x10$^6$ n/cm$^2$/s. Radiographs were recorded of the metal foil sample with an exposure time of 15 s. The distance between the aperture and the detector was 5.5 m resulting in a beam divergence of $\Theta = 1.8 \times 10^{-3}$, which fulfils Eq. 3 for all sample to detector distances. The sample was initially placed at a distance Δ=19 mm from the detector. Radiographs were acquired at increasing sample to detector distances to at every 5 mm until an end position of Δ=189 mm to map the transition from an absorption dominated to a phase contrast dominated regime. A sketch of the setup is shown in Figure 1a.

The collected data were flat field and dark field corrected. Before further analysis, diagonal stripes and a background gradient were removed from the data. The stripes were removed by masking out the regions of Fourier space corresponding to their frequency. The background linear gradient was removed by normalizing to the air signal outside the sample. A radiograph of the metal foil sample after correction is shown in Figure 2a.

Phase retrieval based on the assumption of a monochromatic beam with wavelength given by the mean of the spectrum resulted in the image shown in Figure 2b. Phase retrieval assumes that the sample consists of a homogenous material, but this assumption has been found not to be essential. Beltran *et al.* (2010) demonstrated that the neutron cross section and scattering length can be chosen corresponding to a material of interest, which only locally blurs the boundaries between the material of interest and other materials (Beltran *et al.*, 2010; Beltran *et al.*, 2011). Thus, the neutron cross section and scattering length were calculated from a volume-weighted mean of the constants for Al and Zr, to give results of $\sigma_{eff} = 5.04 \times 10^{-28}$ m$^2$ and $b_{eff} = 6.10 \times 10^{-15}$ m.

### 2.4. White beam neutron imaging of bone samples

#### 2.4.1. Bone samples

Human bone was obtained from the Body Donation Programme at Department of Biomedicine, Health, Aarhus University. Required permission was obtained from the Scientific Ethical Board of the Region of Central Denmark (1-10-72-113-15).

Rods of human femoral bone (Ø = 5 mm) from one individual was drilled so the sample long axis coincided with the load-bearing axis of the bone. Therefore, the vascular canals were mainly oriented parallel to the flow direction in the custom-built aluminium sample holder (Figure 1b). Two bone samples were drilled; one comprised both of normal cortical bone and more porous cortical bone, giving rise to large void spaces, and one comprised of normal cortical bone only, so the only void spaces visible at this resolution are blood vessels. In addition to the Harversian canals cortical bone contains a cellular network with lacunae of 1-10 µm linear dimensions interconnected by canaliculi that are a few hundred nm in diameter, which cannot be observed at the resolution used (Wittig *et al.*, 2022).

**2.4.2. X-ray imaging of bone samples**

Prior to the neutron measurements, the bone samples were investigated by X-ray micro tomography using a Xradia 620 Versa X-ray microscope (ZEISS, Germany) with an accelerating voltage of 50 kV, a power of 4.5 W, and a LE2 (low energy) filter. 3201 projections with an exposure time of 8 s each were collected while rotating the sample 360°, providing an isotropic voxel size of 5.5 µm obtained with an optical magnification of 0.4× and no pixel binning. The source-to-sample distance was 17 mm and the sample-to-detector distance was 88.8 mm. The reconstructions were conducted in the TXM Reconstructor Scout-and-Scan software (v16.1.13038, ZEISS, Germany) using a cone beam adapted filtered back projection (FBP) algorithm, rotated and tilted to match the neutron data using ImageJ (Schneider *et al.*, 2012), and exported to Dragonfly (v2021.3, Object Research Systems (ORS) Inc, Montreal, Canada) for segmentation and rendering.

**2.4.3. Neutron imaging of bone samples**

Propagation-based phase contrast neutron imaging of the bone samples was also performed on the ICON beamline at SINQ, PSI. Using a white beam and a 10 mm aperture, Figure 1c, both 2D radiography and 3D tomography data were collected using 80 s exposure time and an isotropic voxel size of 13.5×13.5×13.5 µm$^3$. The distance between the aperture and the detector was 7.5 m (Figure 1c). The first tomography dataset was collected for the bone sample not containing bigger void spaces without the custom-built aluminium sample holder, thus allowing for a small sample-to-detector distance of Δ=7 mm. A second tomography dataset was collected for the more porous bone sample mounted in a custom-built aluminium sample holder (Figure 1d). Consequently, the sample-to-detector distance was larger than before, Δ=48 mm. While pumping $D_2O$ (Eurisotop) using the lowest possible pump setting (~1 mL/h), the liquid transport was followed by time resolved radiography. Afterwards, with the pump off but still connected, tomography data were collected. Both tomography datasets were obtained with 625 regularly spaced projections covering 0°–360° and an exposure time of 80 s per projection.

Stripes were removed by masking out specific frequency regions in Fourier space similar to the metal sample. The projections were normalized and Paganin single distance phase retrieval was performed based on the assumption of a monochromatic beam with a mean wavelength of 3.1 Å (see below for a discussion of the choice of wavelength in phase retrieval). For this purpose, the neutron cross section and scattering length were estimated as a mean of the constants for the constituents of hydroxyapatite. For ring removal, both horizontal and vertical stripes were removed from the sinograms (Münch *et al.*, 2009) and the resulting projections were reconstructed in MATLAB (R2020b, MathWorks Inc., MA, USA) using the FBP algorithm with a Hann filter. This common reconstruction algorithm was chosen because the goal of the present work was to illustrate the power of the phase retrieval method. The sample containing more porous bone was reconstructed with a projection specific centre-of-rotation correction, as the custom sample holder connected to the pump restricted the rotation. Reconstructions were segmented and the resulting void spaces were rendered in Dragonfly.

## 3. Results

We first explored propagation-based phase contrast imaging using a Be-filtered, or pink, beam on the low absorption metal foil sample. Figure 2a shows a radiograph at a sample-detector distance $\Delta=189$ mm. The sample is seen in the middle of the image. At the bottom, where the sample is thickest, the sample is easily identifiable, and it can be seen that the width of the sample decreases towards the top where it bends to the side. To illustrate the effect of the phase retrieval, the phase retrieved image is compared to the negative logarithm of the normalised image, *i.e.*, to $-\ln(I(x,y)/I_0(x,y))$, where $I_0$ is the white beam image, corresponding to the apparent linear attenuation..

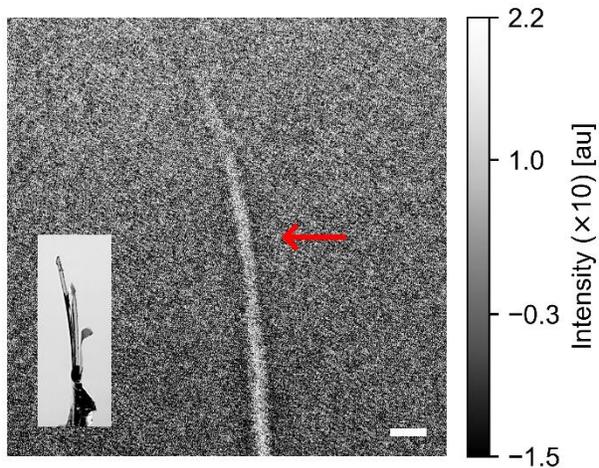

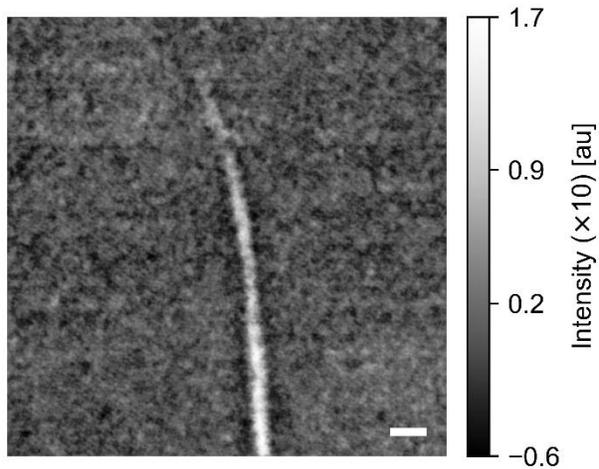

**Figure 2** Neutron radiographs of the metal foil sample, arrow points to sample, at a sample-to-detector distance Δ=189 mm. Scale bars represent 1 mm. Radiographs (corrected as described in the experimental section) a) without phase retrieval and b) after phase retrieval. Inset shows photo of metal foil sample.

Phase retrieval based on the assumption of a monochromatic beam yielded the image in Figure 2b from which a clear improvement in the signal-to-noise ratio is apparent.

To systematically investigate the effect increasing phase contrast we varied the sample to detector distance from near the detector (19 mm) in steps of 5 mm to 189 mm. Figure 3 shows vertical average of 20 rows across the bottom of the sample for every fifth distance measured, corresponding to every 25 mm the sample moved away from the detector.

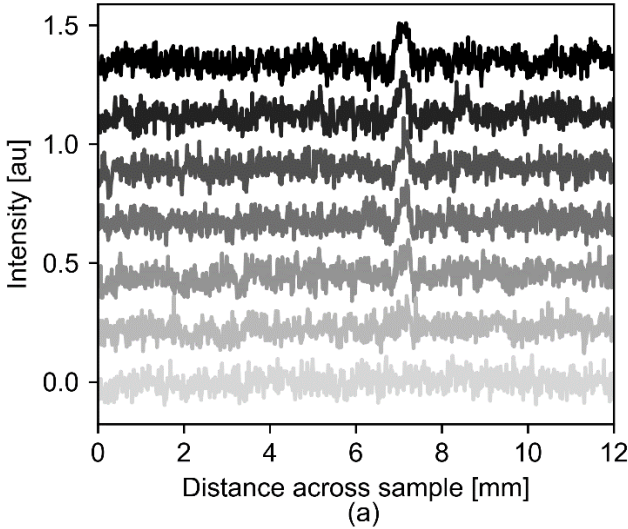

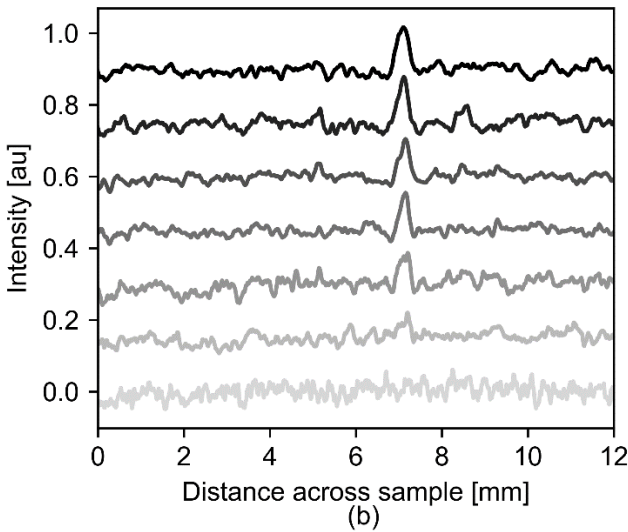

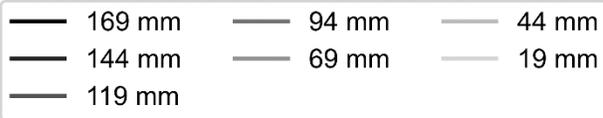

**Figure 3** The phase contrast signal increases for the low absorption contrast metal foil sample when increasing the sample to detector distance. The graphs show a vertical mean over the last 20 lines of the two images shown in Figure 2 where the sample is thickest. Data at every 25 mm sample to detector distance are shown. The graphs are offset vertically for clarity. a) Data without phase retrieval. b) Data after phase retrieval corresponding to Figure 2a and b.

No discernible signal was observed from the sample until a sample to detector distance of Δ=69 mm (Figure 3a). Thus, the sample was effectively 'invisible' at short sample to detector distances due to the low absorption cross section; only upon increasing the distance, and thus increasing phase contrast, the sample became visible. The signal increases linearly with the sample-to-detector distance. This illustrates how propagation-based phase contrast builds up and reveals information almost impossible to detect using absorption contrast. Phase retrieval not only retrieves information but also, and importantly, reduces high frequency noise and thus increases the signal-to-noise ratio (Figure 3b). This clearly demonstrates that a polychromatic neutron beam yields a phase signal. Phase retrieval for this case was performed using the Paganin single distance phase retrieval as defined in Eq. 2 assuming a monochromatic beam and using the average wavelength of the beam spectrum.

In order to compare phase retrieval using the Paganin method (Eq. 2) and the generalised polychromatic version of the Paganin method (Eq. 5), both were applied to the same radiography image (Figure 4a and 4b, respectively). The insets represent the beam spectrum used in the Paganin phase retrieval. For Figure 4a, we assumed a monochromatic beam described by a δ-function with a peak at the average wavelength $\lambda=5.2$ Å while for Figure 4b, we assumed a polychromatic beam with the spectrum shown in the inset. The two results look near identical, as seen from the noise pattern in the image, and comparing line cuts from the radiographies in plot Figure 4d.

This means that functionally there is no difference between using the method defined in Eq. 2 or Eq. 5 for the phase retrieval, as long as the wavelength used in Eq. 2 is the average wavelength of the spectrum. Using a wavelength much higher than the average leads to overdamping and blurring of the signal as seen in Figure 4c and when comparing the line cuts in Figure 4d.

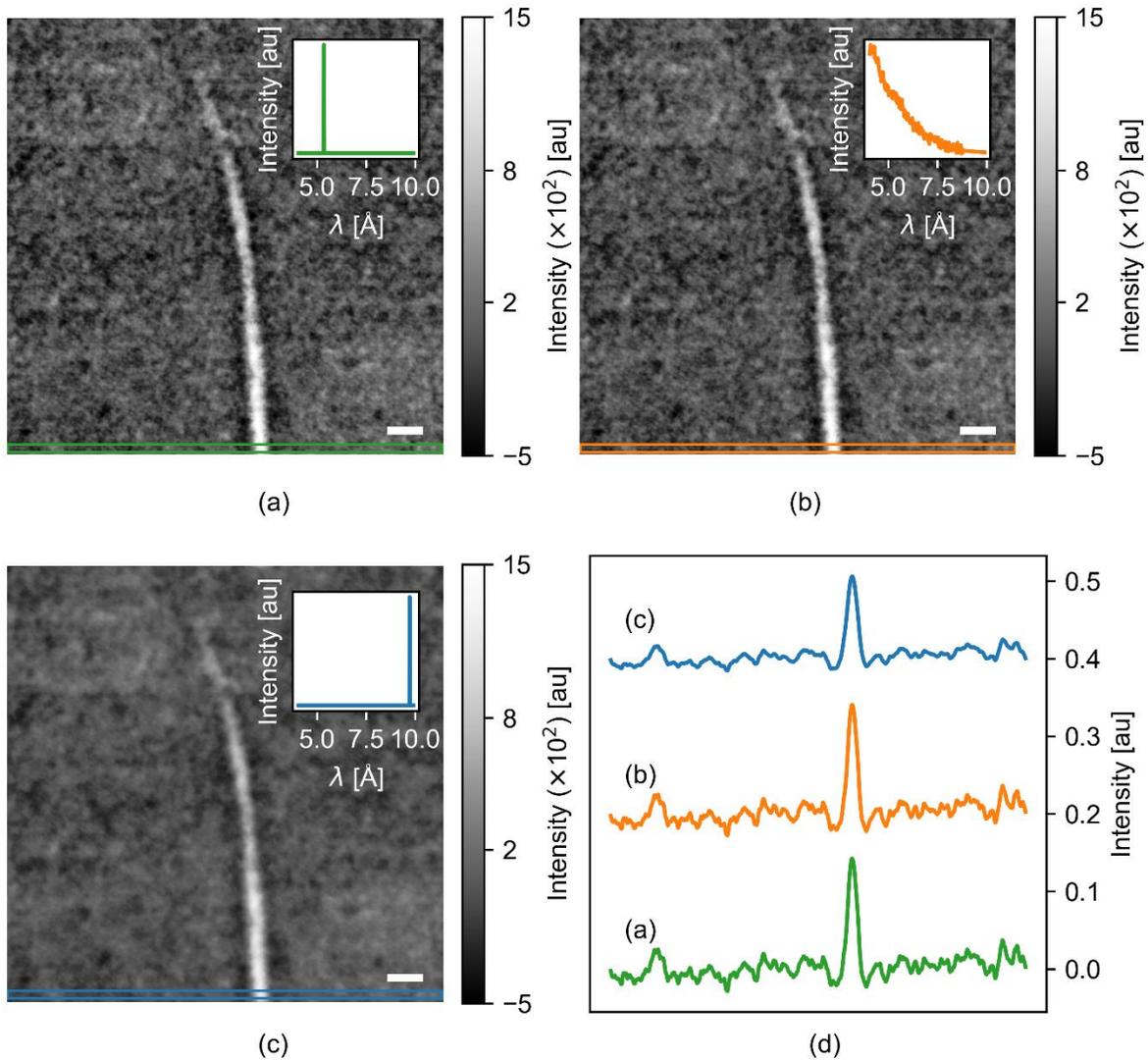

**Figure 4** a), b), and c) Radiograph at a sample to detector distance of Δ=189 mm after energy dependent phase retrieval. The insets show the spectra inserted into the filter. Scale bars represent 1 mm. d) ) Line plot showing a vertical mean over the bottom 20 lines of the images as shown by the coloured boxes in a), b), and c). Data in d) are offset along the *y*-axis for clarity.

### 3.1. Bone results

The bone samples were investigated by X-ray µCT prior to neutron investigations. The samples were then investigated using propagation-based phase contrast neutron tomography, one sample with and the other without $D_2O$ injected into the sample. Figure 5 compares virtual sections through the image stacks and rendering of void spaces from the different experiments. For the neutron tomography, it is clear that, the phase retrieval procedure improves signal to noise considerably, strongly improving the possibility of interpreting the data for example via segmentations similar what was observed for the metal foil sample. We used the X-ray μCT as ground truth for evaluating the ability to detect the vasculature in the samples. The rendered blood vessels are detectable for the phase retrieved neutron data, but less identifiable so for the raw neutron data (Figure 5). Yet, compared to the X-ray renderings, the difference in voxel size and signal-to-noise ratio between the two imaging modalities becomes apparent, since significantly more vascular canals were resolved with X-rays. Nevertheless, the data clearly demonstrate that even with a white beam and a sample with significant absorption, phase retrieval based on propagation-based phase contrast improves signal-to-noise ratio significantly and provides improved ability to identify important structural features in the material. We emphasize that the tomographic reconstructions of the neutron data were performed by simple filtered back projection without any attempts at noise reduction or other efforts to improve reconstruction quality. This was chosen to allow simple evaluation of the impact of phase retrieval for the signal-to-noise of the reconstructed data. The reconstructed neutron data thus do not reflect quality of optimized reconstructions but rather serve to illustrate the improvements attainable by phase retrieval.

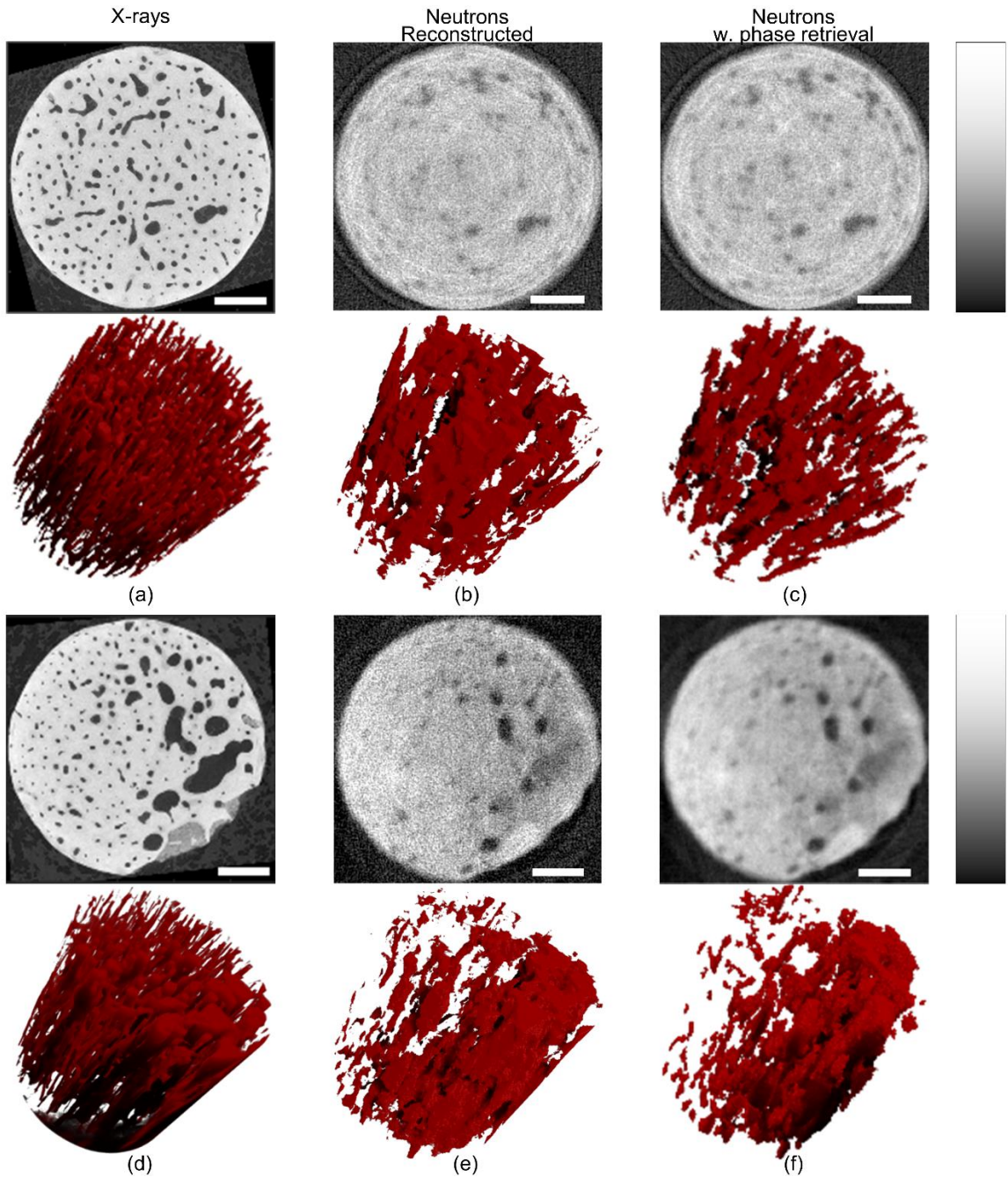

**Figure 5** Bone samples. Typical cross sections and 3D visualisations of the void spaces for (a+d) the lab-based µCT X-ray data, (b+e) the reconstructed neutron data, and (c+f) the phase retrieved neutron data. The top panel show the bone sample comprised of cortical bone only, while the bottom panel show the bone sample comprised of more porous cortical bone. Scale bars represent 1 mm. Corresponding 3D renderings of the segmented blood vessels are shown in red. Images were rotated to be displayed in the same orientation. The contrast in the images is enhanced by limiting the range of the colour scale from 2 –

98% of max values in the data. Colour bars are in the ranges: (a) [635 31127], (b+c) [-0.0009 0.0041], (d) [1244 60940], (e+f) [-0.0003 0.0032].

To evaluate the improvement in signal-to-noise ratio upon phase retrieval we chose a line going through both bone and void space, and used it to compare raw and phase retrieved data (Figure 6). From this, it is evident that the Paganin phase retrieval method improves the signal-to-noise ratio significantly. The Paganin phase retrieval allows for identifying finer features, which is clear from both the line-cuts in Figure 6c and the 3D segmented renderings in Figure 5. In addition, Figure 6d shows grey level histograms from which it is apparent that the noise level in the raw data does not allow for separation of more than air and bone. Nonetheless, additional peaks are visible in the phase retrieved data histogram.

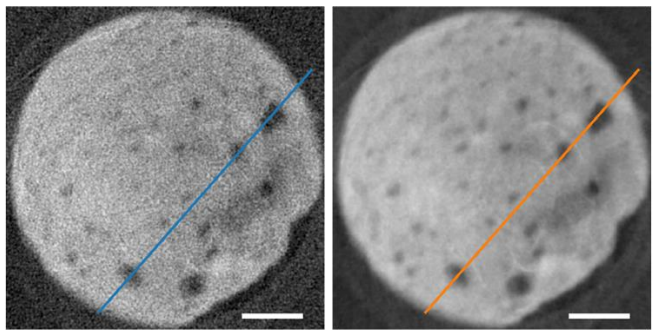

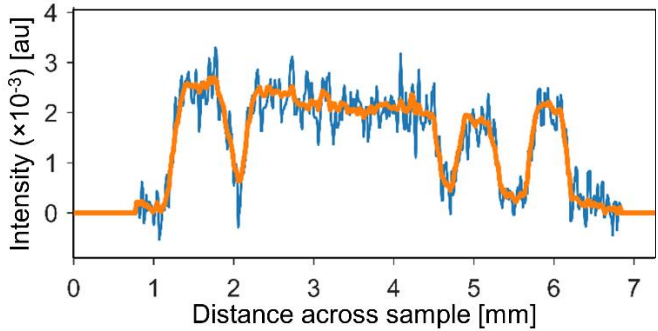

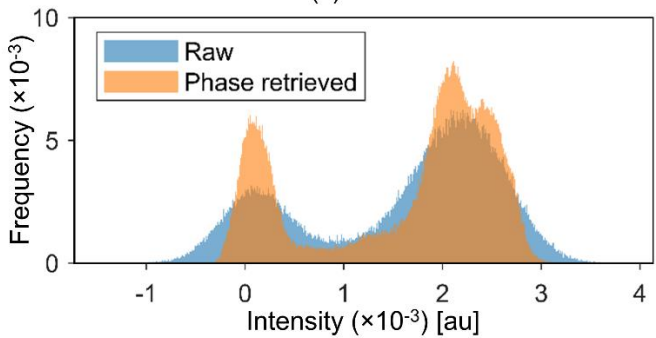

**Figure 6** Lines through reconstructed phase contrast neutron data. (a) Cross section through the reconstructed raw neutron data. The colour bar ranges from -0.0009 to 0.0041. (b) Cross section through the reconstructed phase retrieved neutron data. The colour bar ranges from -0.0009 to 0.0041. Scale bars represent 1 mm. (c) Intensity plot of blue line in (a) and of orange line in (b). (d) Histograms of raw (blue) and phase retrieved (red) neutron data. Bin width is $1 \cdot 10^{-5}$.

The possibility of harnessing propagation-based phase contrast imaging for studying liquid transport was investigated by radiography of the bone sample containing more porous bone. With an exposure time of 80 s, it only took one frame for the $D_2O$ to be seen on the other side of the sample. Due to imperfect sealing at the sample cell/bone interface, some $D_2O$ circumvented the sample as seen by high signals along the edges of the sample (Figure 7a). The sample was purposely oriented with the bigger void spaces to one side and these void spaces light up as white spots when subtracting two detector frames (Figure 7a) indicating presence of $D_2O$ in the bigger bone pore spaces, at least. A neutron propagation-based phase contrast tomogram was collected of the sample while containing $D_2O$. Thus, some of the canals contained $D_2O$. $D_2O$ and bone have quite similar intensity making them difficult to distinguish using absorption neutron imaging (Le Cann *et al.*, 2017; Törnquist, 2021). Indeed, we could not directly segment the water-filled voids from the phase retrieved data. However, correlative imaging combing X-ray and neutron tomograms easily allowed identifying fluid-filled pore spaces. One of the bigger canals/voids is traced with an orange marking in a slice through the X-ray image(Figure 7b). When this region is superimposed to the neutron slice taken after $D_2O$ uptake, the canal is indeed filled with $D_2O$, giving rise to a density closer to bone than to air (Figure 7c). When comparing other canals in these two slices, it is evident that the traced canal is not the only filled with $D_2O$. This illustrates that correlative imaging combining X-ray and neutron contrast can be most helpful in analysing complex materials like bone.

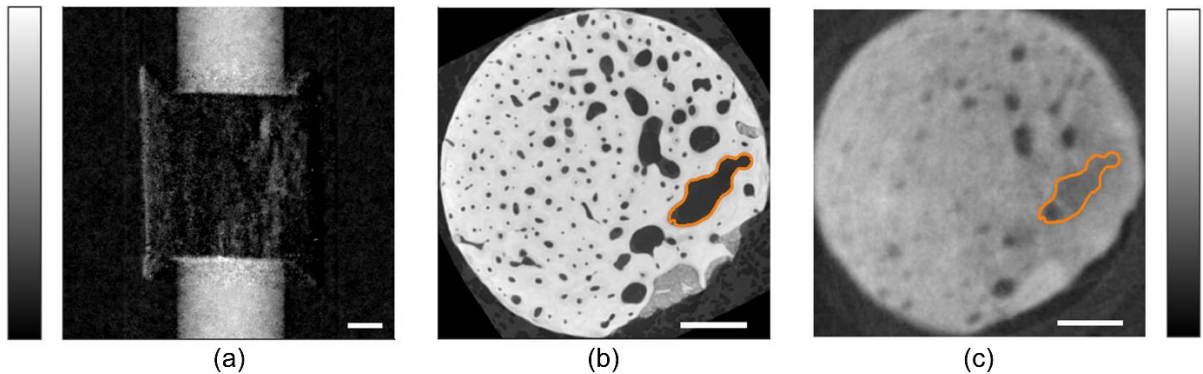

(a)　　　　　　　　　　(b)　　　　　　　　　　(c)

**Figure 7** $D_2O$ uptake in sample containing more porous bone. (a) Two raw neutron detector frames subtracted so that the $D_2O$ lights up on all sides of the sample. White spots mid sample are assigned to $D_2O$ in bigger void spaces, corresponding to area marked with orange in (b) and (c). The colour bar covers the range [0 0.3]. (b) Typical cross section through the lab-based μCT X-ray data with bigger void

marked in orange. Colour bar ranging from [635 31127]. (c) Typical cross section through the phase retrieved neutron data with $D_2O$ filled bigger void marked in orange. The colour bar ranges from -0.0009 to 0.0041. All scale bars equal 1 mm.

To visualise the $D_2O$ in the bone pores, segmentation of a slice of the X-ray data has been superimposed on the neutron dataset of the sample containing $D_2O$, while matching the orientation and rotation of the slices as well as possible. Histograms of the grey level data in the segmented voids and bone are plotted in Figure 8. Three peaks are present in the histogram, one at the same intensity as the bone peak and two at slightly lower intensities, corresponding $D_2O$ and air, respectively. As the resolution of neutron data is lower than for the X-ray data, many of the voids segmented from the X-ray data are not present in the neutron data, giving rise to the relatively large bone peak in the void histogram. The few voids that are present and filled with air in both datasets gives rise to the peak at the lowest intensity in the histogram. The peak at the intermittent intensity therefore corresponds to the $D_2O$ present in the larger void spaces in the sample, indicating a successful penetration of $D_2O$ in the sample.

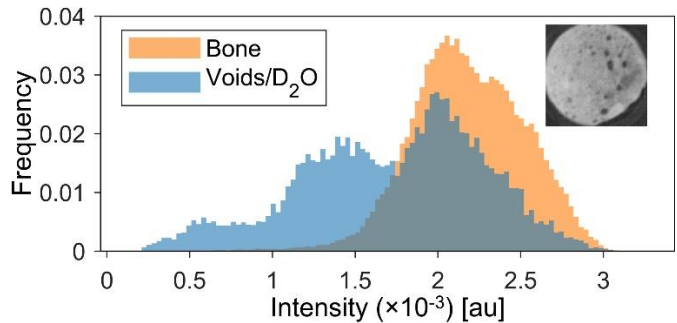

**Figure 8** Histograms of the grey levels of bone in orange and of void spaces identified by X-ray μCT in blue some of which are filled with $D_2O$ in the neutron experiment. Neutron propagation-based phase contrast data of sample containing more porous bone. Bin width is $3 \times 10^{-5}$ a.u.. The inset shows the slice used in the histogram.

## 4. Discussion

Propagation-based phase contrast neutron imaging was recently demonstrated to work on a biological sample using near monochromatic neutrons (Paganin *et al.*, 2019). Here we wish to test if the methodology can be employed also when using polychromatic neutron radiation. We proved the viability of the Paganin phase retrieval method with a polychromatic beam on a metal foil and a bone sample. The results from the metal foil sample showed that the generalised phase retrieval method for a polychromatic beam using a spectrally averaged neutron cross section is functionally no different from using the method for a monochromatic beam, if the wavelength inserted into the method is the weighted mean of the neutron spectrum. This is relevant when imaging materials for which the energy dependent neutron cross section is not known *a priori*.

Both for the metal foil and bone samples, an amplification of signal-to-noise ratio was observed, indicating that phase retrieval can be used to reduce acquisition time, improve image contrast and/or spatial resolution of low absorption materials (Paganin *et al.*, 2019). A potentially larger increase in spatial resolution can be achieved with interferometric methods, which measure both absorption contrast, phase contrast, and dark-field contrast (Pushin *et al.*, 2017; Strobl *et al.*, 2019), but it comes at the expense of an increase in measurement time. This makes neutron interferometry less suited for *in situ* experiments where time resolution is of the essence.

Neutron and X-ray radiation have very different relative scattering cross sections for the elements in the periodic table. Where the X-ray scattering cross section is almost proportional to the electron density, the neutron cross sections fluctuate and are independent of the atomic number. Hence, X-ray and neutron radiation generally lead to very different imaging contrast. Combining the two contrast methods for the same sample therefore allows a more comprehensive study compared to only using one modality as previously discussed by Törnquist *et al.* (2021) and Guillaume *et al.* (2021) and has now been demonstrated in the present study. X-ray radiation provide good contrast between bone, soft tissue, and air. For studies of highly X-ray absorbing metal bone implants, for example, neutrons can be of very good use, since the different contrast avoids metal artefacts stemming from the highly absorbing implant that strongly impedes standard X-ray imaging (Le Cann *et al.*, 2017; Isaksson *et al.*, 2017). Moreover, liquid transport through a piece of bone is another example of the usefulness of neutrons. For this type of experiment, we ideally wish to study a liquid transport through a biological material. With X-rays, it would be necessary to use contrast agents not part of the normal biological environment. With neutrons, $D_2O$ can provide the contrast needed, while being close to the biologic scenario. Secondly, mixtures of $D_2O$ and $H_2O$ can be used to adjust the contrast of the fluid. In the present study, we found that by combining data from X-ray and neutron sources provided more information about the samples than available by either modality on their own. With the phase retrieval algorithm, neutron phase contrast imaging is a good candidate for investigating bone and other low absorption materials where the goal is to distinguish different materials with similar X-ray contrast. Even though we observed that parts of the liquid circumvented the sample due to insufficient sealing between the custom sample holder and the bone sample, the *in-situ* experiment still demonstrated the viability of the phase retrieval approach.

**5. Conclusion**

Propagation-based phase contrast neutron imaging was successfully demonstrated on both a metal sample with very low absorption contrast i.e., an almost pure phase object, and a biological sample, here bone. The results indicate that this method in combination with the Paganin single distance phase retrieval helps decrease noise and increase contrast, which makes it a good method for low attenuation samples and likely to work well with *in situ* experiments. Combining neutron and X-ray radiation for investigating bone and other biological or hierarchical materials provides an increase in complementary information compared to a single modality.

**Acknowledgements**   We acknowledge support from the ESS lighthouse on hard materials in 3D, SOLID, funded by the Danish Agency for Science and Higher Education, grant number 8144-00002B. We thank the Danish Agency for Science, Technology, and Innovation for funding the instrument center DanScatt. Use of the Novo Nordisk Foundation research infrastructure AXIA (grant NNF19OC0055801) and support from Nina K. Wittig is gratefully acknowledged. Research Technician Finn Benthin Saxild, Department of Physics, DTU, is gratefully acknowledged for producing the sample holder.

**References**

Allman, B. E., McMahon, P. J., Nugent, K. A., Paganin, D., Jacobson, D. L., Arif, M. & Werner, S. A. (2000). *Nature* **408**, 158-159.
Alloo, S. J., Paganin, D. M., Morgan, K. S., Gureyev, T. E., Mayo, S. C., Mohammadi, S., Lockie, D., Menk, R. H., Arfelli, F., Zanconati, F., Tromba, G. & Pavlov, K. M. (2022). *Opt. Lett.* **47**, 1945-1948.
Beltran, M. A., Paganin, D. M., Siu, K. K. W., Fouras, A., Hooper, S. B., Reser, D. H. & Kitchen, M. J. (2011). *Phys. Med. Biol.* **56**, 7353-7369.
Beltran, M. A., Paganin, D. M., Uesugi, K. & Kitchen, M. J. (2010). *Opt. Express* **18**, 6423-6436.
Bidola, P., Morgan, K., Willner, M., Fehringer, A., Allner, S., Prade, F., Pfeiffer, F. & Achterhold, K. (2017). *J. Microsc.* **266**, 211-220.
Burger, E. H. & Klein-Nulend, J. (1999). *The FASEB Journal* **13**, S101-S112.
Busse, B., Bale Hrishikesh, A., Zimmermann Elizabeth, A., Panganiban, B., Barth Holly, D., Carriero, A., Vettorazzi, E., Zustin, J., Hahn, M., Ager Joel, W., Püschel, K., Amling, M. & Ritchie Robert, O. (2013). *Sci. Transl. Med.* **5**, 193ra188.
Cowin, S. C. & Cardoso, L. (2015). *J. Biomech.* **48**, 842-854.
Fiori, F., Hilger, A., Kardjilov, N. & Albertini, G. (2006). *Meas Sci Technol* **17**, 2479-2484.
Guillaume, F., Le Cann, S., Tengattini, A., Törnquist, E., Falentin-Daudre, C., Albini Lomami, H., Petit, Y., Isaksson, H. & Haïat, G. (2021). *Phys. Med. Biol.* **66**, 105006.
Isaksson, H., Le Cann, S., Perdikouri, C., Turunen, M. J., Kaestner, A., Tägil, M., Hall, S. A. & Tudisco, E. (2017). *Bone* **103**, 295-301.
Jacobson, D. L., Allman, B. E., McMahon, P. J., Nugent, K. A., Paganin, D., Arif, M. & Werner, S. A. (2004). *Appl. Radiat. Isot.* **61**, 547-550.
Kaestner, A. P., Hartmann, S., Kühne, G., Frei, G., Grünzweig, C., Josic, L., Schmid, F. & Lehmann, E. H. (2011). *NIM-A* **659**, 387-393.
Le Cann, S., Tudisco, E., Perdikouri, C., Belfrage, O., Kaestner, A., Hall, S., Tägil, M. & Isaksson, H. (2017). *J Mech Behav Biomed Mater* **75**, 271-278.
Lehmann, E., Lorenz, K., Steichele, E. & Vontobel, P. (2005). *NIM-A* **542**, 95-99.

McMahon, P. J., Allman, B. E., Jacobson, D. L., Arif, M., Werner, S. A. & Nugent, K. A. (2003). *Phys. Rev. Lett.* **91**, 145502.
Münch, B., Trtik, P., Marone, F. & Stampanoni, M. (2009). *Opt. Express* **17**, 8567-8591.
Paganin, D., Mayo, S. C., Gureyev, T. E., Miller, P. R. & Wilkins, S. W. (2002). *J. Microsc.* **206**, 33-40.
Paganin, D. M., Sales, M., Kadletz, P. M., Kockelmann, W., Beltran, M. A., Poulsen, H. F. & Schmidt, S. (2019). *arXiv* 1909.11186.
Pushin, D. A., Sarenac, D., Hussey, D. S., Miao, H., Arif, M., Cory, D. G., Huber, M. G., Jacobson, D. L., LaManna, J. M., Parker, J. D., Shinohara, T., Ueno, W. & Wen, H. (2017). *Phys Rev. A* **95**, 043637.
Robling, A. G. & Bonewald, L. F. (2020). *Annual Review of Physiology* **82**, 485-506.
Schneider, C. A., Rasband, W. S. & Eliceiri, K. W. (2012). *Nat. Methods* **9**, 671-675.
Strobl, M., Valsecchi, J., Harti, R. P., Trtik, P., Kaestner, A., Gruenzweig, C., Polatidis, E. & Capek, J. (2019). *Sci. Rep.* **9**, 19649.
Törnquist, E. (2021). PhD thesis, Lund University, Sweden.
Törnquist, E., Gentile, L., Prévost, S., Diaz, A., Olsson, U. & Isaksson, H. (2020). *Sci. Rep.* **10**, 14552.
Törnquist, E., Le Cann, S., Tudisco, E., Tengattini, A., Andò, E., Lenoir, N., Hektor, J., Raina, D. B., Tägil, M., Hall, S. A. & Isaksson, H. (2021). *Phys. Med. Biol.* **66**, 135016.
van Tol, A. F., Schemenz, V., Wagermaier, W., Roschger, A., Razi, H., Vitienes, I., Fratzl, P., Willie, B. M. & Weinkamer, R. (2020). *PNAS* **117**, 32251-32259.
Wieland, D. C. F., Krueger, S., Moosmann, J., Distler, T., Weizel, A., Boccaccini, A. R., Seitz, H., Jonitz-Heincke, A. & Bader, R. (2021). *Adv. Eng. Mater.* **23**, 2001188.
Wittig, N. K., Østergaard, M., Palle, J., Christensen, T. E. K., Langdahl, B. L., Rejnmark, L., Hauge, E.-M., Brüel, A., Thomsen, J. S. & Birkedal, H. (2022). *J. Struct. Biol.* **214**, 107822.
Yu, B., Pacureanu, A., Olivier, C., Cloetens, P. & Peyrin, F. (2021). *J. Microsc.* **282**, 30-44.